\documentclass[conference]{IEEEtran}
\IEEEoverridecommandlockouts
\usepackage{cite}
\usepackage{amsmath,amssymb,amsfonts}
\usepackage{algorithmic}
\usepackage{graphicx}
\usepackage{textcomp}
\usepackage{xcolor}
\usepackage{hyperref}
\usepackage{soul}
\usepackage{tabularx}
\graphicspath{{figures/}}

\def\BibTeX{{\rm B\kern-.05em{\sc i\kern-.025em b}\kern-.08em
    T\kern-.1667em\lower.7ex\hbox{E}\kern-.125emX}}

\begin{document}

\title{The xApp Store: A Framework for xApp Onboarding and Deployment in O-RAN}

\author{\IEEEauthorblockN{Philip Rodgers}
\IEEEauthorblockA{\textit{School of Computer Science} \\
\textit{University of Glasgow}\\
Glasgow, UK \\
philip.rodgers@glasgow.ac.uk}
\and
\IEEEauthorblockN{Paul Harvey}
\IEEEauthorblockA{\textit{School of Computer Science} \\
\textit{University of Glasgow}\\
Glasgow, UK \\
paul.harvey@glasgow.ac.uk}
}

\maketitle

\begin{abstract}
5G and beyond mobile telecommunication networks are increasingly embracing software technologies in their operation and control, similar to what has powered the growth of the cloud. This is most recently seen in the radio access network (RAN).  In this new approach, the RAN is increasingly controlled by software applications known as \textit{xApps}, and opens the door to third party development of xApps bringing diversity to the ecosystem, similar to mobile phone apps. This model aligns closely with the controllers in the ITU-T architecture for autonomous networks, and provides a pathway towards autonomous operation in the RAN. Unfortunately, no marketplace to host or supply xApps currently exists. 

This work describes our experiences in leveraging open-source O-RAN implementations to design and develop an xApp store.
\end{abstract}


\section{Introduction}


A 5G network is able to communicate with mobile phones by sending and receiving radio waves. This part of the telephone network is called the radio access network (RAN). Traditionally, the RAN has been deployed as a series of dedicated hardware elements which are strongly associated with the vendors that provided them. More recently, these hardware elements are being increasingly replaced by software virtualisation~\cite{CONDOLUCI201865}, similar to what was seen in the cloud domain.

A key element of this new software-based approach is the introduction of modular applications known as xApps~\cite{hoffman2024xapps}, which are promoted by the Open Radio Access Network (O-RAN) Alliance (\autoref{sec:oran}). These dedicated software components are responsible for monitoring and controlling user-defined functions and operations within the RAN, and have already been identified as key enablers of autonomous network behaviour~\cite{kliks2023itu}. By decoupling control logic from the virtualisation runtime, xApps allow network functionality to be updated independently of the underlying hardware. This architecture also opens the door to third-party development, similar to that of the mobile app ecosystem for smartphones.  We describe the role and characteristics of xApps in more detail in \autoref{sec:xapps}. At present, however, no equivalent of the mobile app store exists to support discovery, onboarding, and deployment of xApps in the O-RAN ecosystem.


This paper describes our experiences in trying to leverage open source O-RAN implementations to develop a prototype xApp Store (\autoref{sec:xapp-store}). The xApp Store proposed in this work not only facilitates third-party innovation and rapid deployment of new functionalities, but also lays the groundwork for automated and self-managing RAN components. To the best of our knowledge, this is the first attempt to create such a store. 
Our objective was to define and implement xApp descriptions, validate the conformance of xApps to their descriptions, and run acceptance testing on an O-RAN platform.


A related body of work explores similar marketplace models for nApps, which are non-real-time applications designed to interface with the 5G core and the Non-RT RIC. Notably, the EVOLVED-5G project developed and validated an nApp marketplace to support vertical integration through standardised APIs and onboarding procedures~\cite{li2025integrating}.  While nApps and xApps operate in different timescales and architectural domains, both efforts highlight a growing need for structured, open ecosystems for third-party applications. We discuss this work more in~\autoref{sec:related_work}.

\section{Open Radio Access Network} \label{sec:oran}



Open Radio Access Network~\cite{oran_whitepapers} is designed to address technical and non-technical challenges in traditional RAN systems.
The high cost of these systems and vendor lock-in is bad for competition and innovation, preventing network operators the flexibility to build cost-effective solutions~\cite{oran_open_ecosystem}.
By fostering a multi-vendor ecosystem, O-RAN aims to reduce costs, allow for third-party innovation, and enable operators to scale and adapt network components to various deployment scenarios~\cite{oran_towards_open_smart_ran}.

At its core, O-RAN is built on two fundamental principles: \textit{vendor-agnostic interoperability} and \textit{open architecture}. The vendor-agnostic nature of O-RAN is achieved through the definition of open interfaces for different tasks, such as RAN control via the E2, A1 or O1 interfaces. These open interfaces enable communication between components from different vendors~\cite{oran_whitepapers}. The intention is that network operators are no longer limited by proprietary ecosystems, and are able to mix and match hardware and software components to suit their needs~\cite{oran_use_cases}. Open interfaces also facilitate the development of standardised service models, such as Key Performance Metrics (KPM) and RAN Control (RC), driving competition and enabling smaller, more specialised, vendors to enter the market~\cite{oran_map}.

\begin{figure}[t]
    \centering
    \includegraphics[width=0.4\textwidth]{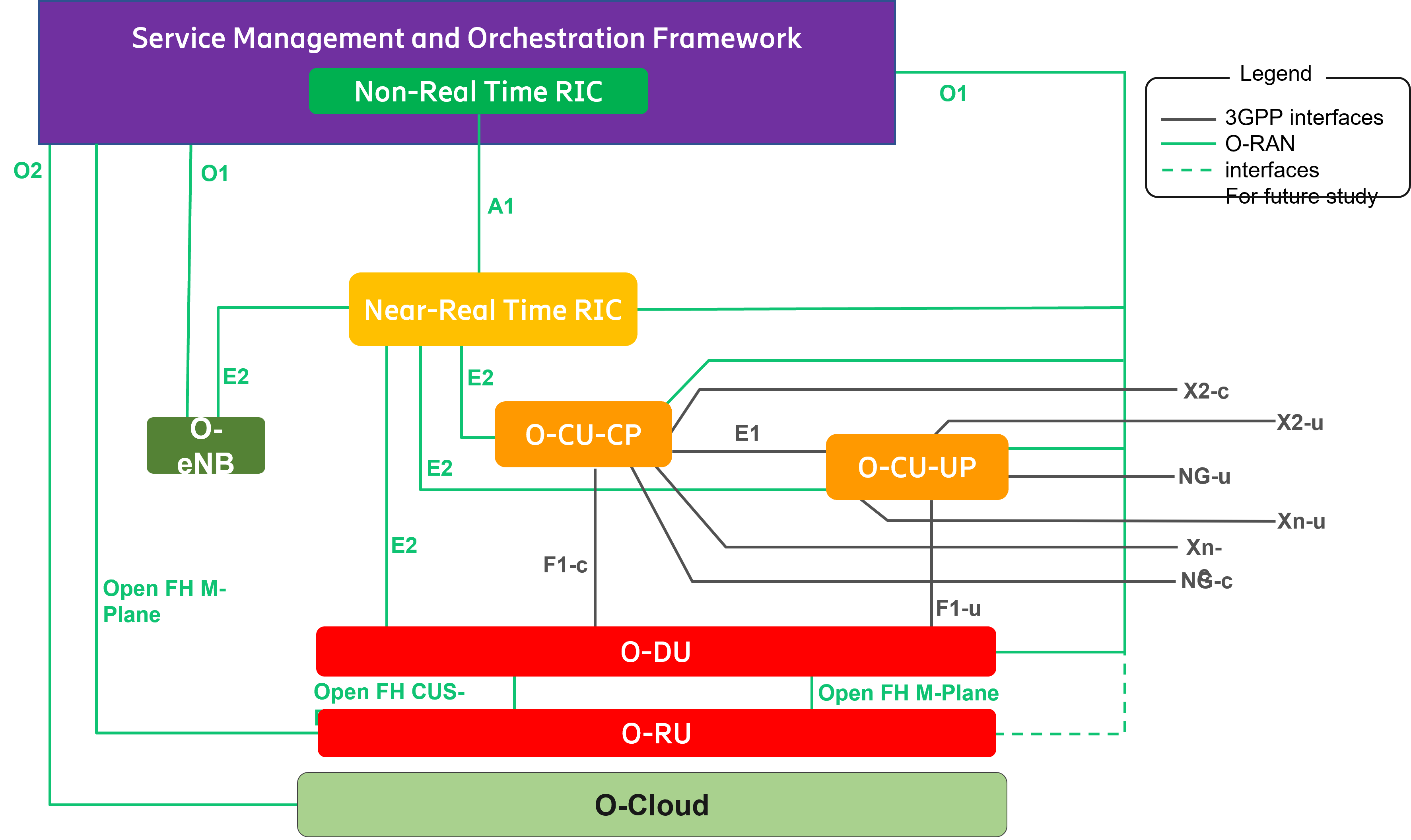}
    \caption{O-RAN Architecture~\cite{o_ran_architecture}}
    \label{fig:o-ran-architecture}
\end{figure}

The open architecture of O-RAN disaggregates the traditional RAN into distinct components, including the Open Radio Unit (O-RU), Open Distributed Unit (O-DU), and Open Central Unit (O-CU)~\cite{oran_minimum_viable_plan, oran_towards_open_smart_ran}. These components communicate through standardised interfaces, allowing for independent upgrades and replacements.
An overview of O-RAN components and interfaces can be seen in \autoref{fig:o-ran-architecture}.


Central to O-RAN’s architecture are the ran intelligent controllers (RIC). There are two variants: the Near-Real-Time RAN Intelligent Controller (Near-RT RIC) and the Non-Real-Time RAN Intelligent Controller (Non-RT RIC)~\cite{niknam2020intelligent}. The Near-RT RIC is intended to support control operations with a timescale of ten milliseconds to one second, enabling time-sensitive control through modular applications called \textit{xApps}~\cite{oran_whitepapers}. In contrast, the Non-RT RIC focuses on long-term network optimisation and analytics, leveraging \textit{rApps} to inform policy and decision making~\cite{oran_minimum_viable_plan}. As our work is focused on xApps we only consider the near-real-time RIC, however, we believe that our approach would be transferable to rApps.  We will refer the near-real-time RIC as RIC in the remainder.


\section{xApps}\label{sec:xapps}
xApps are software applications designed to run on the RIC in O-RAN architectures. They separate control logic from the virtualisation runtime, allowing network functionality to be managed and updated independently of the underlying hardware.  xApp examples include monitoring and control of key network operations such as traffic management~\cite{sroka2024policy}, interference reduction~\cite{yang2024interference}, and performance monitoring~\cite{kouchaki2025performance}. By isolating these functions, network operators can update or replace specific control features without disrupting the entire system.

\begin{figure}[t]
    \centering
    \includegraphics[width=0.4\textwidth]{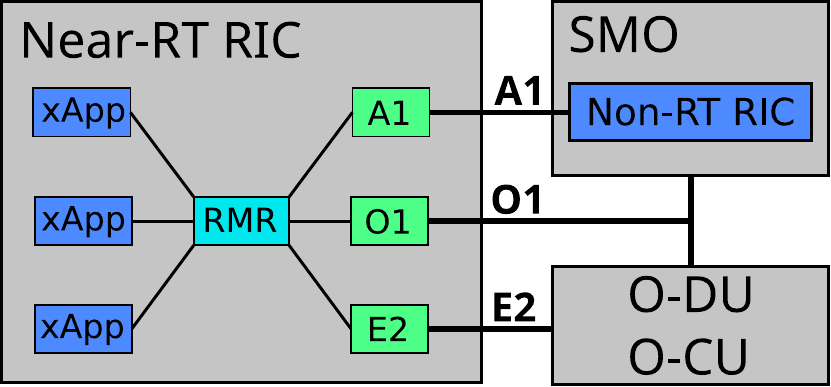}
    \caption{xApps within the near-real-time RIC}
    \label{fig:xapps_in_ric}
\end{figure}

Developed as containerised applications or stand-alone binaries, xApps use standard interfaces (e.g. E2 interface) via RIC Message Router (RMR)\footnote{\url{https://docs.o-ran-sc.org/projects/o-ran-sc-ric-plt-lib-rmr}} to ensure they work across different RIC implementations. When xApps are deployed to the RIC, they register with RMR with the message types it will be sending and receiving, see \autoref{fig:xapps_in_ric}.  While not a publish/subscribe system, RMR routes messages using the message types.  Multiple endpoints can register interest in certain message types so that a single message can be routed to multiple recipients.  This approach is intended to support interoperability and simplify the process of adding or removing network functions as needed.  The xApps' only requirements are to be able to communicate with RMR and be able to formulate and understand the interface messages.  An example workflow would be a traffic steering xApp receiving strategy updates from the Non-Real-time RIC via the A1 interface. It could then send control messages to the radio units via the E2 interface~\cite{rimedolabs_policy_xapp}.

The term ``xApp'' is more than just a label for what an application is intended to do.  It comes with a formal definition as part of the O‑RAN Alliance architecture\footnote{\url{https://docs.o-ran-sc.org/en/latest/}}. As xApps are designed to run on RICs, they are expected to interact with other O‑RAN components using the defined interfaces and protocols. This standardisation is intended to ensure interoperability across the ecosystem.

If an application bypasses the expected messaging framework (RMR) and communicates directly with radio hardware, for example, it is deviating from the formal specifications for xApps.  So while the intended functionality might remain, from a standards and technical standpoint an app that does not adhere to the specified communication methods would no longer be an xApp in the formal sense.

Unfortunately, the current state of the standard interfaces often requires xApp developers to bypass the standard interfaces in order to realise their functionality~\cite{hoffman2024xapps} (\autoref{sec:discussion}).  

\section{xApp Store}
\label{sec:xapp-store}
Our goal for this work was to design and implement a prototype \textit{xApp Store}.  
Similar to a mobile phone app store, the xApp Store is envisaged as a digital distribution platform specifically designed for hosting and managing xApps within an O-RAN ecosystem.

An xApp Store would be a logically centralised repository where developers can publish their xApps and users can browse, download, and install them.  An xApp Store would provide publishing tools that allow developers to upload xApps, manage updates, and monitor performance and usage statistics, managed licensing and payments, and become a marketplace for individual or several network operators.

We believe the creation of an xApp Store is vitally important for O-RAN adoption.  If realised, such a system would enable the deployment and interoperability of xApps in O-RAN by creating a vendor-neutral platform where xApps can be stored, discovered, and deployed across various RIC implementations.  The goal is to provide a standarised description and packaging of xApps, ensuring compatibility across different RIC implementations, and establishing a marketplace for third-party xApp developers. The xApp Store and the xApps it hosts directly map to the knowledge base and controllers of the ITU-T autonomous network architecture~\cite{itu-y3061}, respectively.

\begin{figure}[t]
    \centering
    \includegraphics[width=0.4\textwidth]{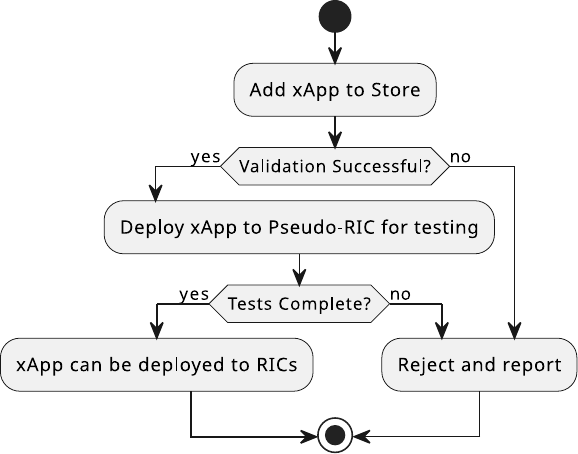}
    \caption{xApp Lifecycle}
    \label{fig:xapp-lifecycle}
\end{figure}

The lifecycle of an xApp in the xApp Store is shown in \autoref{fig:xapp-lifecycle}.

\subsection{Manifest-based validation}

In order to be accepted as valid by the xApp Store, xApps must come bundled with a \textit{manifest file}, similar to the manifest files that come with mobile apps for mobile app stores\footnote{\url{https://developers.google.com/apps-script/concepts/manifests}}.  This file should include information about components, permissions, requirements, and runtime configurations. This manifest file contains the information necessary to achieve automated deployment and validation of xApps within an O-RAN environment.  It allows RIC platforms to verify compatibility and assign necessary resources before running the xApp.  It also allows the RICs to orchestrate deployments based on the interdependence of xApps.

Building on our initial concept~\cite{kliks2023itu}, the manifest includes:

\begin{itemize}
    \item Metadata: Name, version, author, license, and contact information.
    \item Runtime Configuration: Compatible RIC versions and resource requirements.
    \item Interface Definitions: Supported E2 service models and message types.
    \item Dependencies and Security Settings.
\end{itemize}


Some xApps, such as OSC's KPIMON\footnote{\url{https://github.com/o-ran-sc/ric-app-kpimon-go}}, include a JSON file which is used when deploying to a RIC.  This file includes metadata such as author and version, Docker container information and health probes, and messaging information such as which message types the xApp will be sending and receiving.

In its current form, the xApp Store uses the JSON file from the xApp.  This file does not include the all of the features we discussed in our previous work, as we have focused on creating an initial implementation, but this will be extended in the future.  An example of a complete JSON file, based on a combination of our previous work and that of the OSC, is available online and validated against a JSON Schema.\footnote{\url{https://github.com/philrod1/ric-stuff/tree/main/files/manifest}}

\subsection{xApp Testing}

In order to test xApps in an environment we controlled, we had to build our own RIC.  We called this testing environment the \textit{Pseudo-RIC}.  Using the Pseudo-RIC, we can determine if an xApp is deployable and will interact with the RIC components and the simulated RAN components.  We identified and tried four RIC implementation, described more in \autoref{sec:discussion}.  We chose to go with the OSC reference implementation because we believed this would be the most standards compliant, and because the O-RAN Alliance provides a great deal of technical specification documents~\cite{o_ran_specifications}.  Ideally, we would implement all four of the RIC environments outlined in \autoref{sec:discussion} in order to test the xApps in each, but as the xApps we were using to test our system were built for OSC RICs we only implemented an OSC Pseudo-RIC. In this work, a lack of portability of xApps emerged as a key challenge.


Once validated against its manifest, an xApp is deployed to a \textit{Pseudo-RIC} environment.  The behaviour of the xApp is tested using a simulated 5G scenario as described in \autoref{sec:scenario}.  The behaviour is compared to the expected behaviour of the xApp determined by the manifest file.  If an xApp fails validation or testing, a report is generated.  This gives the xApp developer the information required to make their xApp O-RAN compliant.

\subsection{Deployment}
After successful testing on the Pseudo-RIC, xApps become available for deployment to production RICs.  In order to do this automatically, the RIC infrastructure must allow the xApp Store the required access.  This would likely be in the form of a software daemon or agent running on the RIC through which the xApp Store interacts with the RIC.  In our prototype system, this is done via a web API provided by \textit{RICMON}.  RICMON is a web front end which incorporates many RIC tools and is discussed in \autoref{sec:ricmon}.

\section{RICMON}\label{sec:ricmon}
In the course of this work, we discovered that there is a significant amount of fragmentation in the RIC ecosystem, see \autoref{sec:discussion}. Consequently, it was necessary to create a standardised way of interacting with RIC implementations. To this end, we created the \textit{RICMON} tool\footnote{\url{https://github.com/philrod1/ricmon}}. RICMON is a web application built to simplify interactions with a RIC. RICMON is able to provide straight-forward access to the status and logging information from a RIC via REST API calls.  It allows for easy control of RIC functions such as container monitoring and management.


Aside from monitoring the RIC, RICMON's main use is to simplify the deploying and removal of xApps.  To deploy an xApp, RICMON takes a Git repository URL for the xApp.  The URL can be passed to RICMON via its API or its web interface.  It will then \emph{automatically} clone the repository and organise the files into a standardised format.  With all the required files in their correct location, the xApp is \emph{automatically} built using Docker and saved to a locally hosted Docker repository.  We can then use the tools provided by the O-RAN Alliance to onboard and install the xApp to the RIC.

\section{Scenario Simulation}\label{sec:scenario}
Beyond validation against a manifest and deployment of the xApp, the xApp store also seeks runtime verification of the xApps behaviour, and eventually benchmarking. To the best of our knowledge, no standard xApp benchmarking suite exists.

As a first step, a simple mobile network simulation environment was created. The goal of the simulation environment is to test how the xApps interact with the radio systems, such as Next Generation Node B (gNB) and User Equipment (UE) elements they are built to monitor and control. The simulation environment is composed of four parts: -
\begin{itemize}
    \item Logic and mobility simulation, written in JavaScript as an extension of RICMON.
    \item Networking and connectivity implemented using \textit{GNURadio}.
    \item The 5G gNB and UE entities implemented using \textit{srsRAN}.
    \item Network traffic generated using \textit{iperf}.
\end{itemize}

\begin{figure}[t]
    \centering
    \includegraphics[width=0.4\textwidth]{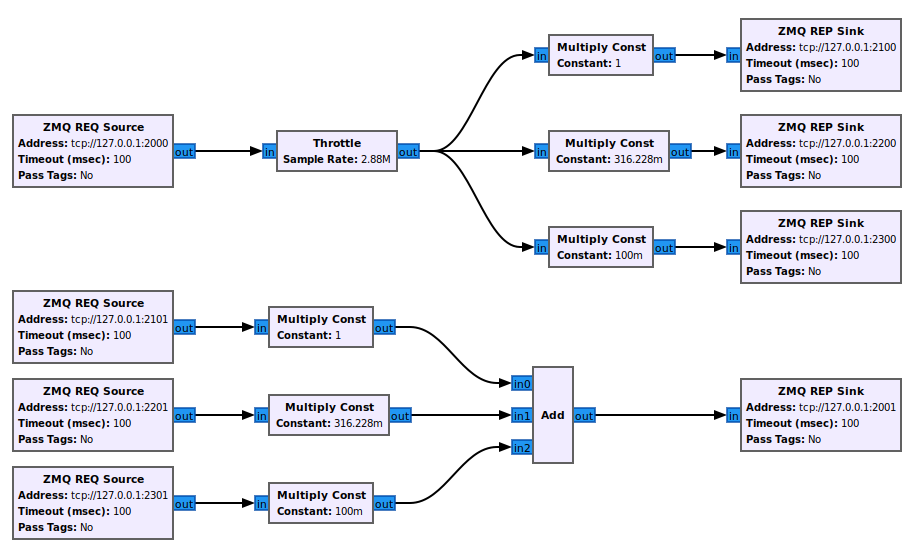}
    \caption{A GNURadio flow chart with three UEs connected to one gNB}
    \label{fig:gnuradio}
\end{figure}

The logic and mobility simulation is responsible for the visualisation and location of the entities.  Each entity is modelled using srsRAN~\cite{flakowski2023srsran} to handle 5G configurations such as frequencies, handover, etc., with network connectivity realised using GNURadio\footnote{\url{https://www.gnuradio.org/}}.  A simplified diagram of GNURadio connectivity can be seen in \autoref{fig:gnuradio}.

We use the simulation to check communication between xApps and the radio equipment they are controlling or monitoring.  At the time of writing, the xApps can monitor the simulation but they cannot control it.  This is due to a combination of lack of support in the form of E2 service models, and also seems to be a limitation of srsRAN software implementations.  The srsRAN documentation shows that real hardware would accept and react to commands sent via E2.  



\section{Discussion}
\label{sec:discussion}
In developing the xApp store it was necessary to deploy a RIC. We now discuss the different options we considered to understand their appropriateness. 

The RIC ecosystem is still evolving, but in this section we focus on four prominent RIC environments from the available literature: OSC RIC, µONOS-RIC, FlexRIC, and OAIC RIC.  Throughout this project we built examples of each of these RIC environments, sometimes multiple examples, with varying levels of difficulty and success.


\subsection{Open-source RICs}

\subsubsection{OSC RIC}\label{sec:osc_ric}
The OSC RIC\footnote{\url{https://docs.o-ran-sc.org/en/latest/}} is the de-facto reference implementation provided by the O-RAN Alliance. It is widely adopted in academic research and initial xApp development due to its adherence to published O-RAN standards. However, the OSC RIC suffers from several practical limitations. Its dependency on legacy software versions (e.g., Ubuntu 20.04 and GCC 9.3.0) and rigid installation scripts pose significant challenges for scalability and integration with more modern tools \cite{polese2023understanding}. Moreover, the standard E2 interface implemented in OSC RIC is often found to be inflexible, limiting the ability to support active control functions such as dynamic traffic steering \cite{santos2024managing}.

\subsubsection{µONOS-RIC}\label{sec:uonos_ric}
µONOS-RIC\footnote{\url{https://docs.onosproject.org/}} is a production-oriented platform designed for large-scale deployments. Its modular design and enhanced scalability make it attractive for industry applications. However, while µONOS-RIC promises improved performance, it tends to deviate from strict adherence to O-RAN standards. This divergence necessitates further testing to fully understand its trade-offs in terms of interoperability and compliance \cite{polese2023understanding, santos2024managing}. As a result, µONOS-RIC may offer better performance in real-world scenarios, but its long-term viability depends on further refinement and industry collaboration. 

\subsubsection{FlexRIC} \label{sec:flexric}
FlexRIC\footnote{\url{https://gitlab.eurecom.fr/mosaic5g/flexric}} is a platform that prioritises low latency and flexibility in xApp interactions. One of its distinguishing features is an extended version of the E2 interface (often referred to as E2*), which permits xApps to bypass certain RIC functions and interact directly with RAN components. This approach reduces latency, making FlexRIC particularly suitable for time-sensitive applications. While this design choice improves performance, it also raises questions about compatibility with other O-RAN-compliant systems, as FlexRIC’s modifications may not be universally supported \cite{chen2023flexapp}.

\subsubsection{OAIC RIC} \label{sec:oaic_ric}
OAIC RIC\footnote{\url{https://openaicellular.github.io/oaic/}}, based on the \textit{E release} of the OSC RIC, provides a middle ground between strict standards compliance and necessary modifications for enhanced functionality. Although it retains a high degree of compatibility with O-RAN specifications, OAIC RIC incorporates adjustments that enable better support for testing environments, such as integration with srsRAN~\cite{} components. Despite these modifications, OAIC RIC is still perceived as less robust compared to production-focused platforms like µONOS-RIC, and it faces challenges similar to OSC RIC in terms of scalability and long-term maintenance \cite{santos2024managing}.

\subsubsection{Summary} In summary, while OSC RIC remains the baseline for academic research due to its standards compliance, its practical limitations have lead to the development of alternative platforms. µONOS-RIC offers potential improvements for production use, FlexRIC provides enhanced low-latency performance through interface extensions, and OAIC RIC represents a compromise with additional support for testing scenarios. The divergence between these academic implementations and production-ready systems underscores the need for further collaboration and refinement of O-RAN standards.  An overview comparison of the four main RIC environments can be seen in \autoref{tab:ric-comparison}.

\subsection{Standards Conformance}

A common observation was that many existing xApps are bypassing the RIC platform entirely for communication with the RAN, instead directly communicating with the E2 interface for performance reasons. This is orthogonal to the RIC design intentions. From the xApp store perspective, this creates a significant challenge in validating xApps if arbitrary communication channels exist between the xApp and the RAN interfaces.  To attempt to evaluate this, we used EdgeShark\footnote{\url{https://edgeshark.siemens.io/}} to sniff network packets being sent in the containerised Pseudo-RIC environment.  Unfortunately, we were unable to observe meaningful behaviour.


\begin{table}[h!]
\centering
\scriptsize
\caption{Comparison of RIC Implementations}
\label{tab:ric-comparison}
\begin{tabularx}{\columnwidth}{|X|c|c|c|c|}
\hline
\textbf{Feature}                & \textbf{OSC RIC} & \textbf{µONOS-RIC} & \textbf{FlexRIC} & \textbf{OAIC} \\
\hline
\textbf{Open-Source}            & \checkmark       & \checkmark         & \checkmark       & \checkmark    \\
\hline                  
\textbf{Standards Compliance}   & \checkmark       & \checkmark         & Partial          & Partial       \\
\hline                  
\textbf{Interface Support}      & E2, A1, O1       & E2, A1, O1         & E2               & E2            \\
\hline                  
\textbf{xApp Modularity}        & \checkmark       & \checkmark         & \checkmark       & \checkmark    \\
\hline                  
\textbf{Production-Ready}       & $\times$         & \checkmark         & $\times$         & $\times$      \\
\hline                  
\textbf{AI/ML Support}          & Partial          & \checkmark         & \checkmark       & \checkmark    \\
\hline                  
\textbf{Good for Research}      & \checkmark       & \checkmark         & \checkmark       & \checkmark    \\
\hline                  
\textbf{Industrial Deployment}  & \checkmark       & \checkmark         & $\times$         & $\times$      \\
\hline                  
\textbf{Ease of Integration}    & Moderate         & High               & Moderate         & High          \\
\hline
\end{tabularx}
\end{table}

The initial tests indicate that a standardised manifest approach simplifies xApp validation and onboarding as the xApps need to provide the necessary information to build, test and deploy them.  The xApp store automates testing and reduces manual steps. However, reliance on older software and the limited capabilities of the current E2 interface hinder full functionality. These issues suggest that while the framework is well-suited for academic research and early-stage development, further enhancements are needed.  Ideally, large telcos who are actively deploying O-RAN solutions will upstream their implementations.

Our investigations reveal several insights and challenges:

\begin{itemize}
    \item \textbf{Integration Challenges:}
    Integrating a reference OSC RIC with modern components such as srsRAN proved to be more complex than anticipated. The OSC RIC’s dependency on older software versions (e.g., Ubuntu 20.04 with GCC 9.3.0) restricted our ability to incorporate up-to-date libraries and tools.  Such issues raise concerns about long-term viability of the reference implementation.
    \item \textbf{xApp Validation and Testing:}
    The introduction of an xApp manifest file and an associated validation process is effective in standardising descriptions and deployment requirements. The xApp Store was able to automate initial validation steps; however, the testing phase on the Pseudo-RIC uncovered limitations in current service model support. Although passive operations (such as KPM monitoring) were successful, active functions (such as traffic steering) could not be fully tested due to limitations of the E2 interface implementation.
    \item \textbf{Toolchain and Ecosystem Constraints:}
    RICMON and the xApp Store provide useful functionality for managing deployments and monitoring states. Yet, our experience underscored the need for a more cohesive ecosystem that minimises manual interventions. The necessity to modify source code for features like dynamic registration of message types (mtypes) highlights a broader issue: the current state of open RAN implementations often requires ad hoc solutions to bridge gaps between evolving standards and existing deployments~\cite{santos2024managing}.
    \item \textbf{Implications for Industry Adoption:}
    The challenges we encountered suggest that while the OSC RIC and associated tools offer a promising testbed for academic research and early-stage xApp development, there may be significant challenges for large-scale, production-level deployments. This dichotomy between research-friendly and production-grade platforms must be addressed to realise the full potential of open, vendor-neutral RAN architectures.
\end{itemize}

\subsection{Recommendations}
Based on our experience, to support future open-source development and engagement of this work we recommend the following: -
\begin{itemize}
    \item Modernising RIC implementations to support current operating systems and toolchains.
    \item Enhancing the E2 interface to better support active functions such as traffic steering.
    \item Expanding testing environments to include real-time scenarios and dynamic message registration.
    \item Fostering industry collaboration to further standardise xApp descriptions and deployment processes.
\end{itemize}

\section{Related Work}\label{sec:related_work}

In the non-real-time domain, the EVOLVED-5G project introduced and validated a comprehensive marketplace for nApps—software modules that interface with the 5G core via standard northbound APIs such as NEF and CAPIF. This marketplace includes mechanisms for onboarding, automated enrolment, and service discovery, as well as deployment support for both operator-managed and third-party domains~\cite{li2025integrating}. A key feature of this approach is the use of semantic descriptors and the TMF Open API framework to expose nApps as catalogue entries.


Additionally, Charpentier et al.~\cite{charpentier2025advancing} provide a broader view of nApps within the 5G and beyond ecosystem. They describe the evolution of plug-and-play interfaces such as NEF, CAPIF, and SEAL, and introduce the concept of a Network Application Package.  This aligns with our manifest-based description for xApps, reinforcing the idea that network programmability should be supported by standard packaging and deployment metadata.  

\section{Conclusions}\label{sec:conclusions}
This work describes the design and implementation of a prototype open-source xApp store to host, test, and deploy O-RAN xApps to support ecosystem development and greater autonomous operation of the network. Leveraging open-source O-RAN implementations, we were able to perform validation, testing and deployment of xApps, noting that our manifest-based approach streamlines the onboarding and testing process.  Simulation results demonstrate that the framework reliably supports passive functions, such as KPI monitoring. Based on our experience, we identified several challenges with the state of the open-source O-RAN ecosystem, which are described and recommendations provided to bridge the gap between academic prototypes and production-ready systems.

\section*{Acknowledgment}
This work was supported by EPSRC IAA award EP/X5257161/1.

\bibliographystyle{IEEEtran} 
\bibliography{references} 

@misc{oran_whitepapers,
  author       = {{O-RAN Alliance}},
  title        = {O-RAN White Papers and Resources},
  howpublished = {\url{https://www.o-ran.org/o-ran-resources}},
  note         = {Accessed: 2025-01-08}
}

@misc{oran_use_cases,
  author       = {{O-RAN Alliance}},
  title        = {O-RAN Use Cases and Deployment Scenarios},
  howpublished = {\url{https://mediastorage.o-ran.org/white-papers/O-RAN.WG1.Use-Cases-and-Deployment-Scenarios-White-Paper-2020-02.pdf}},
  note         = {Accessed: 2025-01-08}
}

@misc{oran_minimum_viable_plan,
  author       = {{O-RAN Alliance}},
  title        = {O-RAN Minimum Viable Plan and the Acceleration towards Commercialization},
  howpublished = {\url{https://mediastorage.o-ran.org/white-papers/O-RAN.Minimum-Viable-Plan-and-Acceleration-towards-Commercialization-white-paper-2021-06.pdf}},
  note         = {Accessed: 2025-01-08}
}

@misc{oran_open_ecosystem,
  author       = {{VIAVI Solutions}},
  title        = {O-RAN: An Open Ecosystem to Power 5G Applications},
  howpublished = {\url{https://www.viavisolutions.com/en-us/literature/o-ran-open-ecosystem-power-5g-applications-white-papers-books-en.pdf}},
  note         = {Accessed: 2025-01-08}
}

@misc{oran_towards_open_smart_ran,
  author       = {{O-RAN Alliance}},
  title        = {O-RAN: Towards an Open and Smart RAN},
  howpublished = {\url{https://assets-global.website-files.com/60b4ffd4ca081979751b5ed2/60e5afb502810a0947b3b9d0_O-RAN%2BWP%2BFInal%2B181017.pdf}},
  note         = {Accessed: 2025-01-08}
}

@misc{oran_map,
  author       = {{O-RAN Alliance}},
  title        = {O-RAN Map},
  howpublished = {\url{https://map.o-ran.org/}},
  note         = {Accessed: 2025-01-08}
}

@article{niknam2020intelligent,
  title        = {Intelligent O-RAN for Beyond 5G and 6G Wireless Networks},
  author       = {Niknam, Solmaz and Roy, Abhishek and Dhillon, Harpreet S and others},
  journal      = {arXiv preprint arXiv:2005.08374},
  year         = {2020},
  url          = {https://arxiv.org/abs/2005.08374}
}

@inproceedings{chen2023flexapp,
  author = {C.-C. Chen and M. Irazabal and C.-Y. Chang and A. Mohammadi and N. Nikaein},
  title = {FlexApp: Flexible and Low-Latency xApp Framework for RAN Intelligent Controller},
  booktitle = {ICC 2023 - IEEE International Conference on Communications},
  address = {Rome, Italy},
  publisher = {IEEE},
  year = {2023},
  pages = {5450--5456},
  doi = {10.1109/ICC45041.2023.10278600}
}

@article{flakowski2023srsran,
  author = {W. Flakowski and M. Krasicki and R. Krenz},
  title = {Implementation of a 4G/5G Base Station Using the srsRAN Software and the USRP Software Radio Module},
  journal = {JTIT},
  volume = {3},
  number = {2023},
  pages = {30--40},
  year = {2023},
  doi = {10.26636/jtit.2023.3.1298}
}

@article{santos2024managing,
  author = {J. F. Santos and A. Huff and D. Campos and K. V. Cardoso and C. B. Both and L. A. DaSilva},
  title = {Managing O-RAN Networks: xApp Development from Zero to Hero},
  journal = {arXiv preprint},
  year = {2024},
  note = {arXiv:2407.09619},
  doi = {10.48550/arXiv.2407.09619}
}

@misc{o_ran_architecture,
  author       = "{O-RAN Software Community}",
  title        = "{O-RAN Architecture Documentation}",
  year         = "2025",
  url          = "https://docs.o-ran-sc.org/en/j-release/architecture/architecture.html",
  note         = "Accessed: 2025-01-21"
}

@misc{o_ran_specifications,
  author       = "{O-RAN Alliance}",
  title        = "{O-RAN Specifications}",
  year         = "2025",
  url          = "https://specifications.o-ran.org/specifications",
  note         = "Accessed: 2025-01-21"
}

@ARTICLE{polese2023understanding,
  author={Polese, Michele and Bonati, Leonardo and D’Oro, Salvatore and Basagni, Stefano and Melodia, Tommaso},
  journal={IEEE Communications Surveys and Tutorials}, 
  title={Understanding O-RAN: Architecture, Interfaces, Algorithms, Security, and Research Challenges}, 
  year={2023},
  volume={25},
  number={2},
  pages={1376-1411},
  doi={10.1109/COMST.2023.3239220}}

@article{CONDOLUCI201865,
title = {Softwarization and virtualization in 5G mobile networks: Benefits, trends and challenges},
journal = {Computer Networks},
volume = {146},
pages = {65-84},
year = {2018},
issn = {1389-1286},
doi = {https://doi.org/10.1016/j.comnet.2018.09.005},
url = {https://www.sciencedirect.com/science/article/pii/S1389128618302500},
author = {Massimo Condoluci and Toktam Mahmoodi}
}

@ARTICLE{hoffman2024xapps,
  author={Hoffmann, Marcin and Janji, Salim and Samorzewski, Adam and Kulacz, Lukasz and Adamczyk, Cezary and Dryjański, Marcin and Kryszkiewicz, Pawel and Kliks, Adrian and Bogucka, Hanna},
  journal={IEEE Journal on Selected Areas in Communications}, 
  title={Open RAN xApps Design and Evaluation: Lessons Learnt and Identified Challenges}, 
  year={2024},
  volume={42},
  number={2},
  pages={473-486},
  doi={10.1109/JSAC.2023.3336190}}

@ARTICLE{kliks2023itu,
  author={KLIKS, Adrian and DRYJANSKI,  Marcin and Ram OV, Vishnu  and Wong, Leon and Harvey, Paul },
  journal={ITU Journal on Future and Evolving Technologies}, 
  title={Towards autonomous open radio access networks},
  year={2023},
  volume={4},
  number={2},
  pages={251-268},
  doi={https://doi.org/10.52953/GJII3746}}

@techreport{itu-y3061,
author = {ITU-T},
type = {Standard},
key = {ITU Y.3061},
month = dec,
year = {2023},
title = {{Autonomous Networks -Architecture Framework}},
volume = {2000},
address = {Geneva, CH},
institution = {United Nations International Telecommunication Union}
}

@ARTICLE{sroka2024policy,
  author={Sroka, Pawel and Kulacz, Lukasz and Janji, Salim and Dryjański, Marcin and Kliks, Adrian},
  journal={IEEE Transactions on Vehicular Technology}, 
  title={Policy-Based Traffic Steering and Load Balancing in O-RAN-Based Vehicle-to-Network Communications}, 
  year={2024},
  volume={73},
  number={7},
  pages={9356-9369},
  doi={10.1109/TVT.2024.3399924}}

@INPROCEEDINGS{yang2024interference,
  author={Yang, Hanchao and Mcpeak, Connor and Nagampally, Rakesh and Tripathi, Nishith and Anderson, Gustave and Reed, Jeffrey H. and Yang, Yaling and Jakubisin, Daniel J.},
  booktitle={MILCOM 2024 - 2024 IEEE Military Communications Conference (MILCOM)}, 
  title={Agile 5G Networks: Advance Traffic Steering xAPP for Interference Mitigation}, 
  year={2024},
  volume={},
  number={},
  pages={300-305},
  doi={10.1109/MILCOM61039.2024.10773839},
  ISSN={2155-7586},
  month={Oct},}

@ARTICLE{kouchaki2025performance,
  author={Kouchaki, Mohammadreza and Natanzi, Seyed Bagher Hashemi and Zhang, Minglong and Tang, Bo and Marojevic, Vuk},
  journal={IEEE Communications Magazine}, 
  title={O-RAN Performance Analyzer: Platform Design, Development, and Deployment}, 
  year={2025},
  volume={63},
  number={2},
  pages={152-159},
  doi={10.1109/MCOM.002.2300822},
  ISSN={1558-1896},
  month={February},}

@misc{rimedolabs_policy_xapp,
  title        = {Policy-based Traffic Steering: xApp Implementation within O-RAN},
  author       = {{RimeDo Labs}},
  howpublished = {\url{https://rimedolabs.com/blog/policy-based-traffic-steering-xapp-implementation-within-o-ran/}},
  note         = {Accessed: 2025-03-06; n.d.},
}

@ARTICLE{li2025integrating,
  author={Li, Xi and Mesodiakaki, Agapi and Gatzianas, Marios and Kalfas, George and Charismiadis, Anastasios-Stavros and Tsolkas, Dimitris and Zanzi, Lanfranco and Salkintzis, Apostolis and Warren, Dan and Gavrielides, Andreas and Sophocleous, Marios and Gramaglia, Marco and Gavras, Anastasius and Cogalan, Tezcan and Lee, Haeyoung and Bulakci, Omer},
  journal={IEEE Communications Magazine}, 
  title={Integrating nApps in 5G for Verticals: Architecture Innovation and Technology Enablers}, 
  year={2025},
  volume={63},
  number={1},
  pages={161-167},
  doi={10.1109/MCOM.001.2300803}}

@ARTICLE{charpentier2025advancing,
  author={Charpentier, Vincent and Landi, Giada and Giannopoulou, Eleni and Brenes, Juan and Camelo, Miguel and Marquez-Barja, Johann M. and Slamnik-Kriještorac, Nina},
  journal={IEEE Network}, 
  title={Advancing Vertical Services for 6G: Future Directions and Innovations}, 
  year={2025},
  volume={},
  number={},
  pages={1-1},
  doi={10.1109/MNET.2025.3550929}}

\end{document}